\documentclass[aps,prl,reprint,superscriptaddress]{revtex4-2}

\usepackage{graphicx}
\usepackage{amsmath}
\usepackage{amssymb}
\usepackage{float}
\usepackage{bigints}
\usepackage[normalem]{ulem}
\usepackage[colorlinks,urlcolor=blue]{hyperref}
\usepackage[capitalise]{cleveref}
\usepackage{xcolor}

\newcommand{\SM}[1]{{\color{black}{#1}}}

\begin{document}  
    \title{\SM{Multiple Mechanisms for Emerging Conductance Plateaus in Fractional Quantum Hall} States}

	\author{Sourav Manna}
 \email{sourav.manna@weizmann.ac.il}
	\affiliation{Department of Condensed Matter Physics, Weizmann Institute of Science, Rehovot 7610001, Israel}
	\affiliation{Raymond and Beverly Sackler School of Physics and Astronomy, Tel-Aviv University, Tel Aviv, 6997801, Israel}
	
	\author{Ankur Das}
         \email{ankur@labs.iisertirupati.ac.in}
	\affiliation{Department of Condensed Matter Physics, Weizmann Institute of Science, Rehovot 7610001, Israel}
        \affiliation{Department of Physics, Indian Institute of Science Education and Research (IISER) Tirupati, Tirupati 517619, India}

 	\author{Yuval Gefen}
	\affiliation{Department of Condensed Matter Physics, Weizmann Institute of Science, Rehovot 7610001, Israel}

	\author{Moshe Goldstein}
	\affiliation{Raymond and Beverly Sackler School of Physics and Astronomy, Tel-Aviv University, Tel Aviv, 6997801, Israel}
 
\begin{abstract}

\SM{Two-terminal conductance quantization in the context of quantum Hall (QH) physics is intimately related to the current carried by a discrete number of chiral edge modes. Upon pinching off a QH bar, one may engineer setups where some modes are fully transmitted (while the others are fully reflected), giving rise to the orthodox theory of quantized conductance plateaus. Here, we note that the 
observation of quantized plateaus \emph{does not} uniquely indicate the underlying mechanism. Our study demonstrates explicitly that (i) such plateaus may be the manifestations of entirely different mechanisms; (ii) conductance measurements alone will not suffice to 
distinguish one from the other. We further show that measurements of shot noise (auto- and cross-correlation) at the plateau may discriminate among different mechanisms. While our observations 
apply to a broad class of QH states, we demonstrate their applicability employing a prototypical example: the bulk state of filling factor $\nu=2/3$. We present distinctly different  scenarios that lead to a conductance plateau $G_{2-\text{terminal}} = e^2/3h$ (observed previously), and likewise qualitatively different mechanisms 
leading to $e^2/2h$ (recently observed). We also predict the possibility of a new conductance plateau at $5e^2/9h$, following a non-orthodox scenario.}

\end{abstract}	
\maketitle

\textit{Introduction.---} 
The oldest known examples of topological states of matter are quantum Hall states in a two-dimensional electron gas (2DEG)
subject to a strong magnetic field
\cite{PhysRevLett.45.494,PhysRevLett.49.405,RevModPhys.82.3045}. These gapped bulk phases have chiral edge modes \cite{PhysRevD.13.3398,Volkov1985} carrying both charge and energy 
\cite{PhysRevB.25.2185,PhysRevB.41.12838,PhysRevB.43.11025}.
In simple cases, such as the Laughlin states \cite{PhysRevLett.50.1395}, these modes co-propagate. 
The situation 
becomes more intriguing when 
counter-propagating modes appear, either due to topological
constraints \cite{PhysRevLett.64.220,Kane1994} (for hole-like bulk filling, e.g.\ $\nu=2/3$ \cite{PhysRevLett.64.220,Wen1990,McD1991,West1992,Kane1994}) or due to edge reconstruction \cite{PhysRevB.49.8227,Wang2013}.
The edge modes can be unequilibrated \cite{Kane1994,Wang2013} or experience different 
degree of charge and thermal equilibration \cite{PhysRevB.101.075308,Protopopov2017}, giving rise to a whole zoo of scenarios for a given
bulk topological order --- a challenging problem to resolve.

\SM{An essential component for manipulating and controlling edge modes is a quantum point contact (QPC) 
\cite{PhysRevLett.60.848,10.1063/1.881503,PhysRevLett.102.106403}, a 
constriction in 2DEG.
By the orthodox theory, quantized conductance plateaus
appear when some of the edge modes are fully backscattered, while the others are fully transmitted in a partially pinched QPC.
Earlier works generalized this paradigm of 
conductance plateaus to the case of edge reconstruction \cite{Wang2013,Khanna2021,PhysRevLett.129.146801}, edge-mode renormalization \cite{Kane1994,Protopopov2017},
and edge equilibration \cite{PhysRevB.101.075308}. For example,  for $\nu$ = 2/3 bulk filling, plateaus at  $2e^2/3h$ \cite{Protopopov2017,Nosiglia2018,Mahalu2009}, and  $e^2/3h$ \cite{Mahalu2009} have been  predicted \cite{Wang2013} 
and  observed experimentally.}

\begin{figure}[!ht]
\includegraphics[width=0.95\columnwidth]{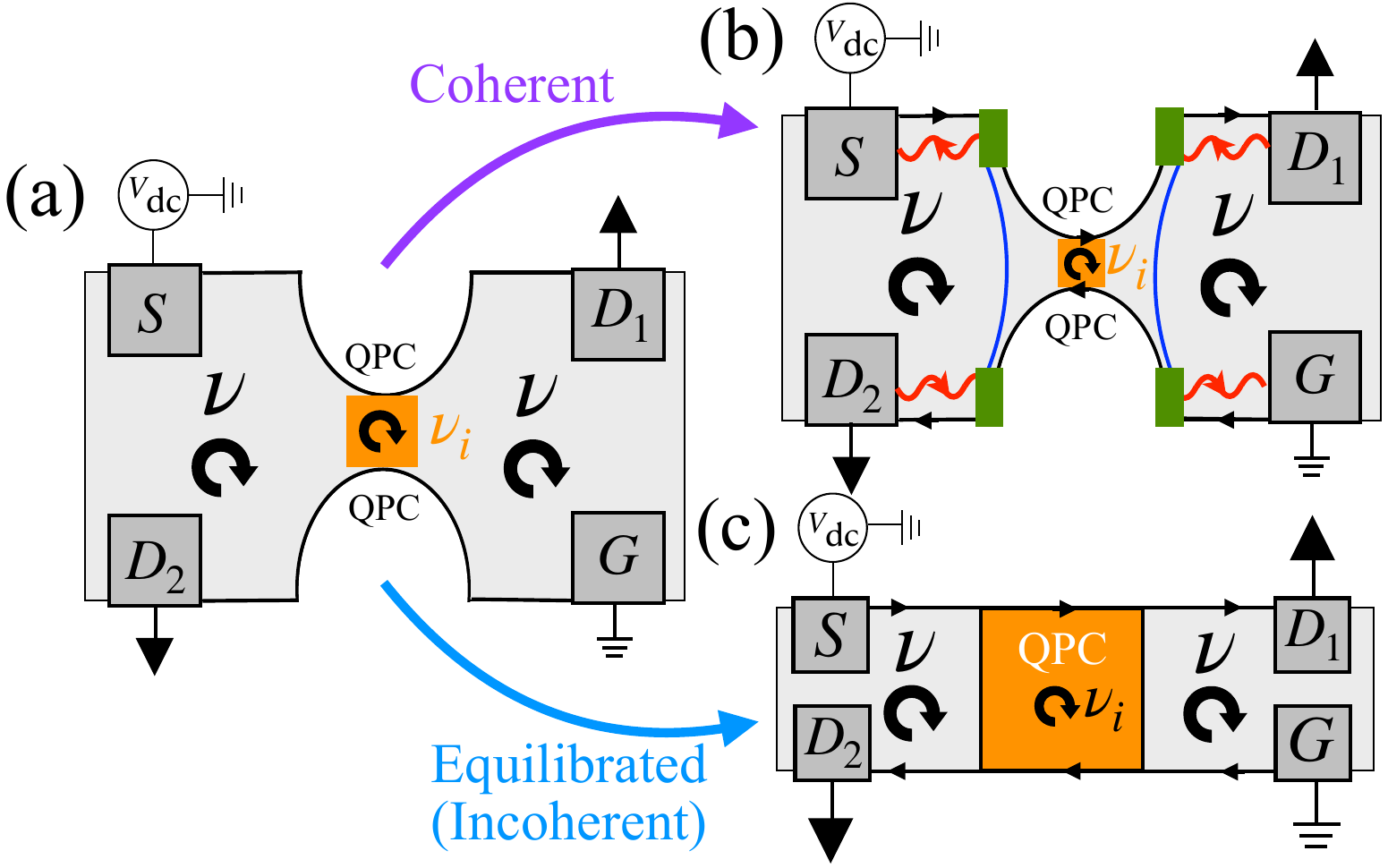}
\caption{
\SM{(a) A Hall bar at filling $\nu$ is interrupted by a
QPC with filling $\nu_i$. Circular arrows indicate
chirality. The contacts include a source $S$ at voltage
$V_{\text{dc}}$, a ground $G$, and two drains $D_1, D_2$.
Each segment of the device can either be in coherent
regime or equilibrated (incoherent), giving rise to the same QPC conductance plateau. (b) Coherent regime — In each segment, the edge modes are renormalized due 
to electron-electron interactions and impurity scattering, leading to decoupled charge (solid black and blue) and neutral modes (wiggly red).
Stochastic/deterministic charge scattering processes occur in the green regions. (c) Incoherent regime — In each segment, the edge modes equilibrate by exchanging charge and energy through electronic interactions and disorder mediated tunneling. An effective hydrodynamic mode (solid black) thus emerges at each segment, characterized by its charge and
thermal conductances.}
}\label{KeyFigure}
\end{figure}

\SM{In a QPC device,
we will have an intermediate small region that can have a different filling 
$\nu_i$ from the bulk filling $\nu$ (c.f.\ \cref{KeyFigure}).
Then, the edge modes at each segment can be in
a distinct steady state—either coherent or equilibrated.
In the coherent regime at zero temperature, interactions and impurity-mediated scattering
lead to a disorder-dominated renormalization group (RG) fixed point, with renormalized charge and neutral modes. These modes take part in coherent transport and give rise to a QPC conductance
plateau. Shot noise can appear at this plateau due to
stochastic charge scattering processes.
In the equilibrated (incoherent) regime at finite temperature, the edge modes interact and exchange charge and
energy due to disorder-induced random tunneling, leading to 
the formation of a hydrodynamic chiral mode, characterized by effective charge and thermal conductances, thus giving rise to a QPC conductance
plateau. Shot noise can appear at this plateau due to
the creation of particle-hole pairs and their stochastic 
splitting \cite{PhysRevB.101.075308,Biswas2022,Khanna2021}. Remarkably, we find that these two fundamentally different mechanisms (coherent vs.\ incoherent)
can give rise to the same emergent QPC conductance plateau for the same bulk state. As we point out, to detect the underlying mechanism, one may use instead a qualitative inequality among the auto- and cross-correlation electric current noise \footnote{\SM{Note that in our previous study \cite{SM5by2} equilibration was assumed, and shot noise was rather used to distinguish between different bulk states at $\nu=5/2$.}}.

Our analysis is
applicable for both 2DEG and graphene in generic quantum Hall states.
We exemplify it for the $\nu=2/3$
state taking edge reconstruction into account. We show that both coherent and incoherent regimes can give rise to 
each of the $e^2/3h$ (observed previously \cite{Mahalu2009}) and 
$e^2/2h$ (recently observed \cite{Manfra2022,Hirayama2023}) conductance plateaus and that
indeed, shot noise serves as the diagnostics \cite{DCG}.
Interestingly, the possibility of a new conductance plateau at $5e^2/9h$ also appears only in the coherent regime.}

\begin{figure}[!t]
	\includegraphics[width=0.99\columnwidth]{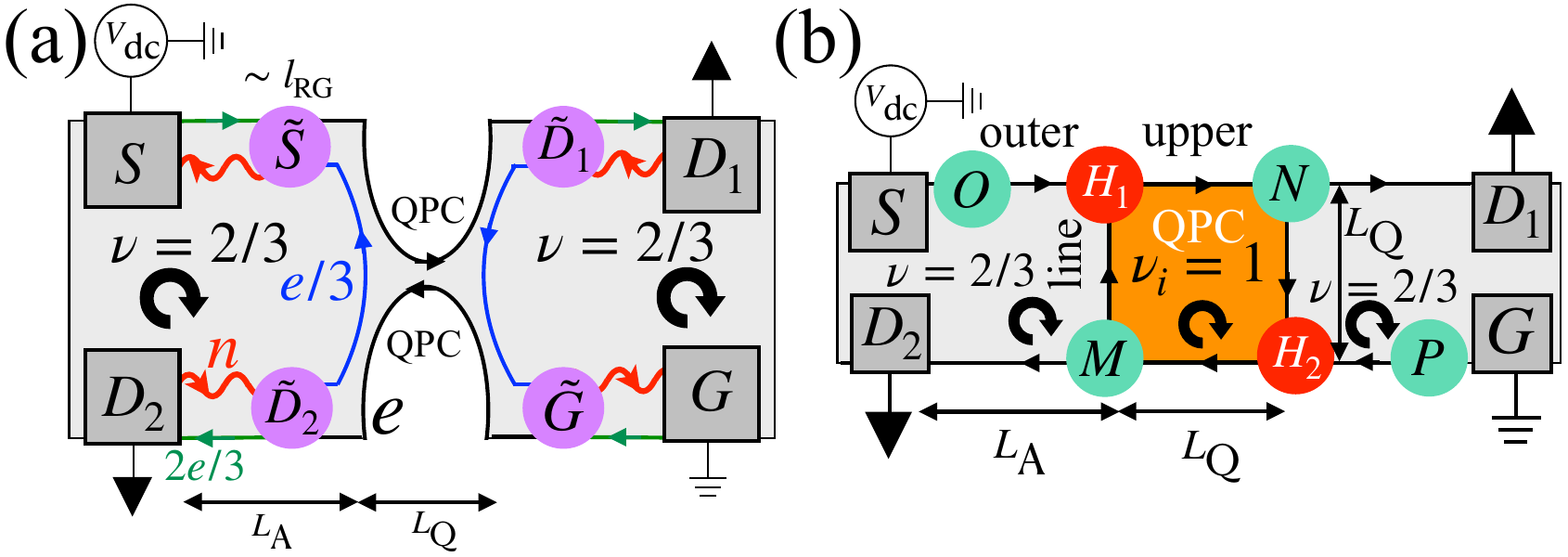}
	\caption{Different scenarios for $G_{D_1}=1/2$; For device geometry, see \cref{KeyFigure}. (a) In the unequilibrated scenario, the
 MacDonald edge structure \cite{PhysRevLett.64.220,Wen1990,McD1991,West1992} renormalizes
 to a counter-propagating $2e/3$ charge mode and a neutral mode $n$ (KFP RG fixed point \cite{Kane1994}) emanating from the contacts. At the QPC, the $e/3\ (e)$ mode is fully backscattered (transmitted). Wavepacket tunneling occurs in the
regions $\Tilde{S}, \Tilde{D}_1, \Tilde{G}, \Tilde{D}_2$
of size $l_{\text{RG}}$.
The charge scattering processes are
stochastic (deterministic) at $\tilde{D}_1$, $\tilde{D}_2$ ($\tilde{S}$, $\tilde{G}$).
(b) Equilibration model with QPC filling $\nu_i=1$. We denote the boundary of the 
vacuum with $\nu$ as the ``outer", and with $\nu_i$ as the ``upper", and between $\nu$ and $\nu_i$ as the ``line". 
Voltage drops occur at the hot spots $H_1,H_2$ resulting in the noise spots $M,N,O,P$ \cite{PhysRevB.101.075308,SM5by2}.} \label{1by2QPC}
\end{figure}

\begin{figure}[!t]
	\includegraphics[width=0.99\columnwidth]{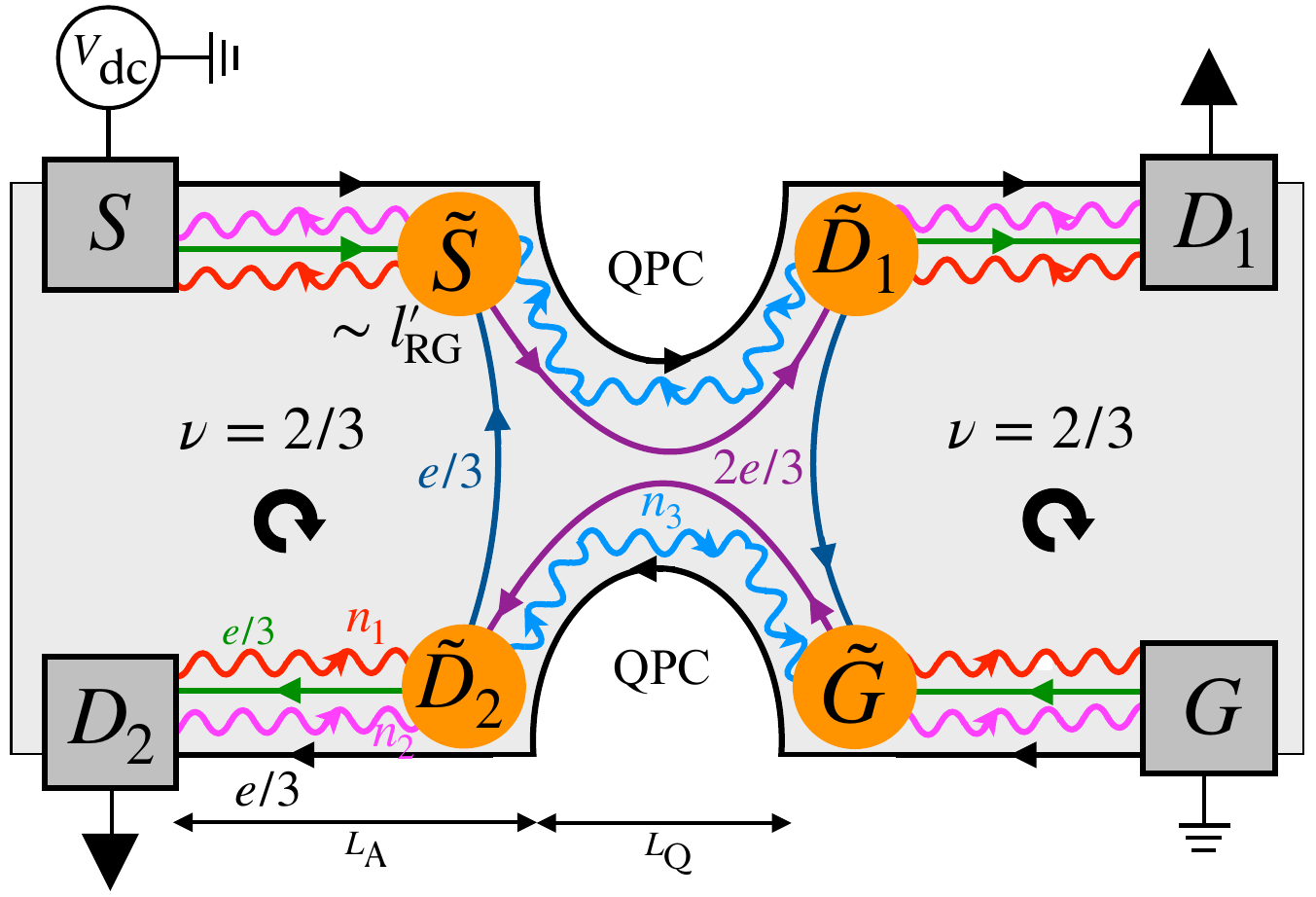}
	\caption{A scenario for $G_{D_1}=5/9$ (c.f.\ \cref{1by2QPC}); In this unequilibrated
 scenario, the reconstructed MacDonald edge
	structure \cite{PhysRevLett.72.2624, Wang2013} renormalizes to a $e/3$ charge mode counter-propagating to the  neutral modes $n_1, n_2$ at the WMG RG fixed point \cite{Wang2013} emanating from the contacts.
 At the QPC, the outermost (innermost) $e/3$ mode is fully transmitted (backscattered) and the remaining modes are renormalized to counter-propagating $2e/3$ charge mode and the neutral mode $n_3$ at the KFP RG fixed point \cite{Kane1994}. 
 Wavepacket tunneling occurs similarly as in \cref{1by2QPC}(a).} \label{5by9QPC}
\end{figure}

\begin{figure}[!t]
	\includegraphics[width=0.99\columnwidth]{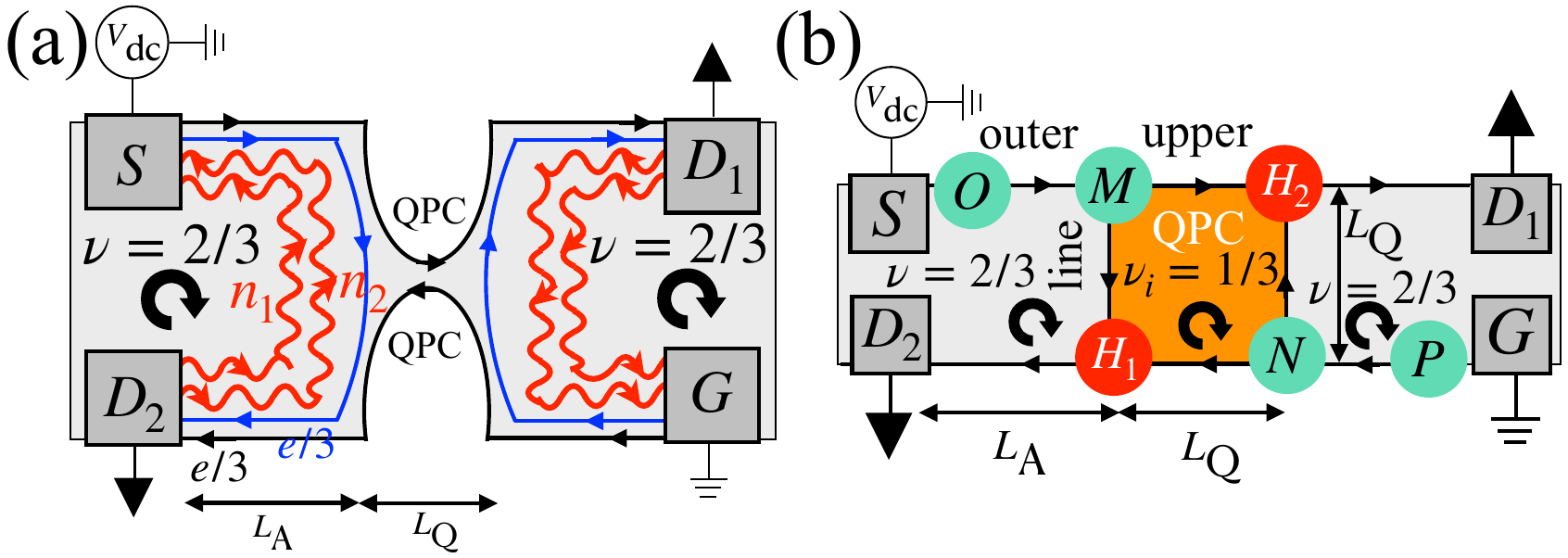}
	\caption{Different scenarios for $G_{D_1}=1/3$ (c.f.\ \cref{1by2QPC}); (a) The renormalized (at the WMG RG fixed point \cite{PhysRevLett.72.2624, Wang2013}) reconstructed MacDonald edge
	structure, consisting of $n_1,n_2\ (\text{neutrals}),e/3\ (\text{inner charge}),e/3\ (\text{outer charge})$ modes is shown. At QPC, the 
 inner (outer) $e/3$ charge mode is fully backscattered (transmitted). (b) Edge equilibration model for $\nu=2/3$ and QPC filling $\nu_i=1/3$ \cite{PhysRevB.101.075308}.} \label{1by3QPC}
\end{figure}

\textit{Setup}.--- 
We assume the typical arm length $L_\text{A}$ $\gg$ the typical QPC size $L_\text{Q}$ (Figs.\ \ref{1by2QPC}, \ref{5by9QPC}, \ref{1by3QPC}).
In addition to the edge modes, dictated by topology, there can be
edge reconstruction leading to the introduction of counter-propagating
edge modes for each filling.
For each edge structure the modes can be in the unequilibrated regime
at zero temperature and can be renormalized
to a RG fixed point or can exhibit equilibration at a finite temperature. 

Recent experiments have
shown that the charge equilibration length $l_{\text{eq}}^{\text{ch}}$
is typically very short
\cite{PhysRevLett.126.216803, Melcer2022, Kumar2022, Srivastav2022}, permitting
full charge equilibration
in each segment of the device: 
$l_{\text{eq}}^{\text{ch}} \ll L_{\text{Q}} \ll L_{\text{A}}$. The thermal equilibration length $l^\text{th}_\text{eq}$ can be parametrically larger, allowing for three regimes of
thermal equilibration: (1) each segment is thermally
unequilibrated,
$L_{\text{Q}} \ll L_{\text{A}} \ll l_{\text{eq}}^{\text{th}}$ (no),
(2) the QPC is thermally
unequilibrated while the other 
segments are thermally equilibrated,
$L_{\text{Q}} \ll l_{\text{eq}}^{\text{th}} \ll L_{\text{A}}$ (hybrid),
and (3) each segment is thermally equilibrated,
$l_{\text{eq}}^{\text{th}} \ll L_{\text{Q}} \ll L_{\text{A}}$ (full).
For full equilibration, the modes 
in each segment form a chiral hydrodynamic
mode characterized by its electrical and thermal conductances, which 
eliminates any effect of edge reconstruction.

$I_1$ and $I_2$ are the currents
(correspondingly $Q_1$ and $Q_2$ 
are the charges)
entering the drains $D_1$ and $D_2$, respectively. 
The dc current-current auto-correlations are defined as $\delta^2 I_1=\langle (I_1 - \langle I_1 \rangle)^2
\rangle$ in $D_1$ and $\delta^2 I_2=\langle (I_2 - \langle I_2 \rangle)^2 \rangle$ in $D_2$, while the cross-correlation is $\delta^2 I_c=\left\langle \left(I_1 - \langle I_1 \rangle\right)
\left(I_2 - \langle I_2 \rangle\right) \right\rangle$ \cite{averagedef}.
Correspondingly, the correlations in charge fluctuations are 
$\delta^2 Q_1 = \langle Q_1^2 \rangle - \langle Q_1 \rangle^2$, $\delta^2 Q_2 = \langle Q_2^2 \rangle - \langle Q_2 \rangle^2$, and $\delta^2 Q_c= \langle Q_1 Q_2 \rangle - \langle Q_1 \rangle \langle Q_2 \rangle$ \cite{SM}.
The Fano factors are defined as $F_j = |\delta^2 I_j|/2e \langle I \rangle t(1-t) = |\delta^2 Q_j|/e \tau \langle I \rangle t(1-t)$, with $j \in \{1,2,c\}$, where $I$ is the source 
current, $\tau$ is time, and $t = \langle I_1 \rangle/\langle I \rangle$ is the QPC transmission \cite{SM}. The QPC conductance is $G_{D_1}e^2/h$, where $G_{D_1}=t\langle I \rangle \tau/e$. 

We focus on the prototypical example of bulk filling $\nu=2/3$ \cite{PhysRevLett.64.220,Wen1990,McD1991,West1992,Kane1994}, its QPC (at filling $\nu_i$)
conductance plateaus, and shot noise in those plateaus.
We show that the auto- and cross-correlation may become 
\emph{unequal} at a plateau, discerning both the edge configuration and its degree of equilibration (\cref{CCC_Table}) assuming
no bulk-leakage
\cite{PhysRevLett.123.137701, Banerjee2018, PhysRevB.99.041302,PhysRevLett.125.157702}.

\begin{table*}
\centering
	\includegraphics[width=\textwidth]{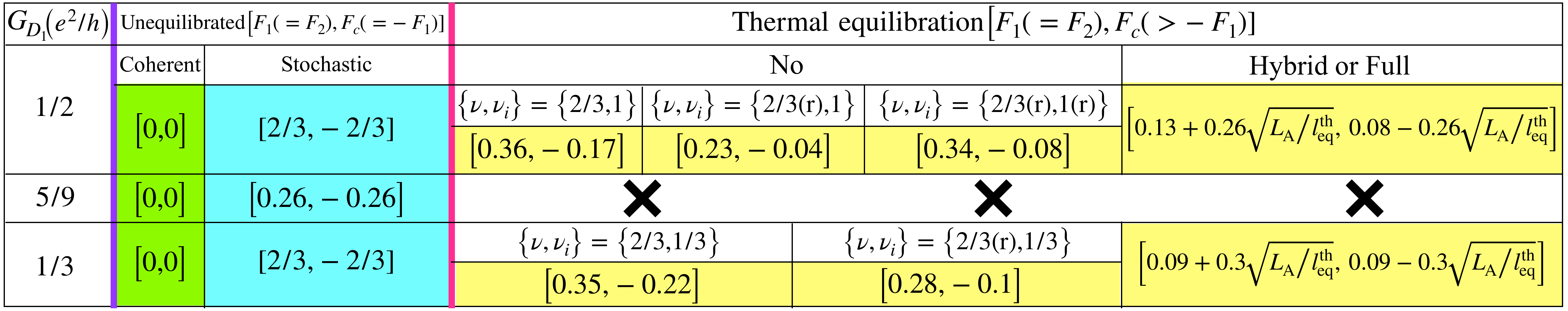}
	\caption{Summary of our results: $G_{D_1}$ represents different QPC (filling $\nu_i$ for $\nu=2/3$) conductance plateaus \cite{Manfra2022,Hirayama2023, Mahalu2009}, and corresponding Fano factors are $F_1, F_2$ (auto-correlations for drains $D_1, D_2$, respectively) and $F_c$ (cross-correlation). 
 $l^{\text{th}}_{\text{eq}}$ is thermal equilibration length and $L_\text{A}$ is arm length (c.f.\ \cref{1by2QPC}).
 Here, $2/3(\text{r})$ refers to the reconstructed MacDonald edge \cite{PhysRevLett.72.2624, Wang2013} and
$1(\text{r})$ denotes edge reconstruction in the QPC filling \cite{Khanna2021}. The effect of edge reconstruction is washed out for the hybrid and full equilibrations.}\label{CCC_Table}
\end{table*}

\textit{The $G_{D_1}=1/2$ plateau.---} 
Recent experiments have shown the emergence of $G_{D_1}=1/2$ \cite{Manfra2022,Hirayama2023} and, similar to an earlier work \cite{PhysRevB.87.125130}, a theoretical explanation was provided \cite{Manfra2022} (see also Refs.\ \onlinecite{Wang2021, Yuli2023}). Here, we show that this plateau may arise due to different mechanisms in either the unequilibrated or equilibrated
regimes and shot noise can be used to discriminate among
them (\cref{CCC_Table}, \cref{1by2QPC}).

(a) Unequilibrated scenario.--- We consider three distinct cases: (A) Under the contacts the modes are clean and noninteracting \cite{Protopopov2017, PhysRevB.104.115416}. Following Ref.\ \onlinecite{Protopopov2017}, we find 
no QPC conductance plateau and no noise in the coherent regime \cite{SM}, leading to a $4e^2/3h$ two-terminal conductance for a fully open QPC.
If in this case one considers a model of incoherent stochastic scattering of charge in transition regions between different segments (as in (C) below), there would be noise even without the QPC, which is incompatible with the observations. 
(B) The $2e/3$ and $n$ modes, at the Kane-Fisher-Polchinski (KFP) RG fixed point, emanate from the 
contacts and undergo coherent scattering at the QPC. Following Ref.\ \onlinecite{Protopopov2017}, we find 
a $G_{D_1}=1/2$ QPC conductance plateau and no noise \cite{SM}.
(C) The $2e/3$ and $n$ modes emanate from the 
contacts and undergo incoherent stochastic scattering in
$\Tilde{D}_1, \Tilde{D}_2$ (while
at $\Tilde{S}, \Tilde{G}$
charge is deterministically transmitted into the single outgoing mode) near the QPC within a region of size RG length 
$l_{\text{RG}}$ (\cref{1by2QPC}(a)).
We consider a wavepacket from $S$ in the mode $2e/3$ in time
$\tau$. It encounters an
infinite number of stochastic (deterministic) reflections and transmissions while entering
and leaving the KFP regions \cite{Kane1994} in $\Tilde{D}_1, \Tilde{D}_2\ (\Tilde{S}, \Tilde{G})$ and leaves the device through either $D_1$ or $D_2$.
The values of the respective reflection and transmission coefficients
are parametrized by the elements
of the density kernel matrix \cite{PhysRevB.52.R17040} in each KFP region. 
Let us denote the outer (inner) mode
as mode-1 (mode-2). Then, the $ij$ element of the density kernel (scattering) matrix at $\tilde{S}$ describes scattering from mode-$i$ to mode-$j$ etc., so one has
\begin{equation}
    S_{\Tilde{S}}=
\begin{pmatrix}
T_{11}  & R_{21} \\
R_{12} &  T_{22}
\end{pmatrix};
S_{\Tilde{D}_1}=
\begin{pmatrix}
\langle T'_{11} \rangle & \langle R'_{21} \rangle\\
\langle R'_{12} \rangle & \langle T'_{22} \rangle 
\end{pmatrix}.
\end{equation}
Similarly, $S_{\Tilde{G}}\ (S_{\Tilde{D}_2})$ has elements 
$T'', R''\ (\langle T''' \rangle, \langle R''' \rangle)$.
Here $T',T''',R',R'''$ are Bernouli random numbers $\in \{0,1\}$, with $\langle \cdots \rangle$ being the first moment of
the distribution, while $T,T'',R,R''$ are deterministic \cite{SM}. 
To calculate the total charge reaching
different contacts we write down an infinite
series of terms, each of which is composed of the following factors: (i) a tunnelling factor for the first entrance to the QPC region from $S$,
(ii) a factor for the shortest path leaving the QPC region to reach a contact ($D_1$ or $D_2$),
and
(iii) a factor giving the contribution of multiple
reflections from different KFP regions; the latter factor is the same for both $D_1$ and $D_2$ \cite{SM}.
This process
gives rise to the shot noise and summing up this series we find
\begin{equation}
\begin{split}
&\langle Q_1 \rangle = \frac{2e}{3} \Big[  \frac{T_{11} \langle T'_{11} \rangle}{1-\langle R'_{12} \rangle R''_{21} \langle  R'''_{12} \rangle R_{21} } \Big],
 \end{split}
\end{equation}
where $T, R$ are transmission and reflection coefficients between the two modes \cite{SM}. Similarly, we find $\langle Q_2 \rangle$ and 
use $T_{11} =1, \langle T'_{11} \rangle=2/3, \langle R'_{12} \rangle=\langle R'''_{12} \rangle=1/3, R''_{21} = R_{21} =1$ leading to $\langle I \rangle = 2e/3\tau$,
$t = 3/4$,
and $G_{D_1} = 1/2$ \cite{SM}. Moreover, $\delta^2 Q_1=\langle Q_1 \rangle(2e/3-\langle Q_1 \rangle)$
and thereby $F_1 = F_2 = -F_c = 2/3$ \cite{SM}.
We note that $G_{D_1}=1/2$ matches with the recent experimental observations \cite{Manfra2022,Hirayama2023} in (B,C).

(b) Equilibration scenario.--- 
We assume that the charge transport is ballistic (B), moving ``downstream" along each segment of the device (\cref{1by2QPC}(b)).
There heat transport can be either B or diffusive (D) or antiballistic (AB, i.e., ``upstream'')
leading to an exponentially suppressed, an algebraically decaying, or a constant shot noise, respectively, as a function of the geometric length of the segment \cite{PhysRevLett.123.137701,PhysRevB.101.075308}.
Voltage drops occur at the hot spots ($H_1, H_2$) leading to the power dissipation
\begin{equation}
	\begin{split}
		P_{H_1} = P_{H_2} = \frac{e^2 V_{\text{dc}}^2}{h}\frac{\nu(\nu_i-\nu)}{2\nu_i} \Big( \frac{\nu_i}{2 \nu_i - \nu}  \Big)^2.
	\end{split}
\end{equation}
This gives rise to the noise spots ($M,N,O,P$), which are formed due to the creation of thermally excited particle-hole pairs
and their stochastic splitting into $D_1, D_2$ (\cref{1by2QPC}(b)) \cite{PhysRevB.101.075308, SM5by2}.
The auto-correlation Fano factors acquire contributions from
$M, N, O, P$, which we denote as $F_M, F_N, F_O, F_P$, respectively, while and the
cross-correlation Fano factor acquire contributions from
$O, P$, to be denoted as $F'_O, F'_P$, respectively, leading to $F_1 = F_2 = F_O + F_P + F_M + F_N$ and $F_c = F'_O + F'_P - F_M - F_N$ \cite{SM}.
Note that, while contribution from
$O, P$ are non-zero, we have \emph{unequal} auto- and cross-correlation. We explicitly write \cite{SM}
\begin{equation}
	\begin{split}
	&2e I t(1-t)(F_M+F_N)=2\frac{e^2}{h}\frac{\nu \nu_i (\nu_i-\nu)}{(\nu-2\nu_i)^2}k_{\text{B}}(T_M+T_N),\\&
	2e I t(1-t)(F_O+F_P)=\frac{1}{(\nu-2\nu_i)^2}\left(\nu_i^2 \mathcal{S}_O + (\nu-\nu_i)^2\mathcal{S}_P\right),\\&
	2e I t(1-t)(F'_O+F'_P)=\frac{\nu_i(\nu_i-\nu)}{(\nu-2\nu_i)^2}(\mathcal{S}_O+\mathcal{S}_P),
	\end{split}
\end{equation}
where $k_{\text{B}}$ is the Boltzmann constant
and $T_M=T_N$ denote the $M, N$ noise spot temperatures \cite{PhysRevB.101.075308,SM5by2,SM}. The piece
$\mathcal{S}_O=\mathcal{S}_P$ is found by evaluating
\cite{PhysRevB.101.075308,SM5by2,SM}
\begin{equation}
	\begin{split}
		&\mathcal{S}_O=\mathcal{S}_P=\\&\frac{2e^2}{hl_{\text{eq}}^{\text{ch}}}\frac{\nu'\nu_{-}}{\nu_{+}} \bigintss_0^{L_{\text{A}}} \frac{e^{\frac{-2x}{l_{\text{eq}}^{\text{ch}}}}k_{\text{B}}\big[T_{+}(x)+T_{-}(x)\big]}{\left[1-\left(e^{\frac{-L_{\text{Arm}}}{l_{\text{eq}}^{\text{ch}}}}\frac{\nu_{-}}{\nu_{+}}\right)\right]^2}dx,
	\end{split}
\end{equation}
where the downstream (upstream) mode in the outer segment has filling $\nu_{+}~(\nu_{-})$ and temperature
profile $T_{+}(x)~(T_{-}(x))$ and $\nu'=\nu_{+}-\nu_{-}$.

Full charge equilibration leads to $G_{D_1} = 1/2, t=3/4$
for three possible edge 
structures (\cref{CCC_Table}).
For no thermal equilibration (Ref.\ \onlinecite{Kumar2022} studied a single edge) we have only B and AB heat transports leading to 
constant Fano factors (\cref{CCC_Table}). 
Here at $O$($P$) only the upstream modes from
$H_1$($H_2$) will contribute to the noise, while the downstream modes from the contacts (at zero temperature) will not contribute. We assume that at the noise spot, all the modes produce noise through the creation of particle-hole pairs. These processes are efficient enough in 
transfering heat energy among the modes to locally equilibrate all the modes to a common temperature
at the noise spot (c.f. \cref{1by2QPC})
leading to constant noise \cite{SM}. For hybrid and full
thermal equilibration, the heat transport in the outer segment becomes D, leading to a 
$\sqrt{L_{\text{A}}/l_{\text{eq}}^{\text{th}}}$ dependence in
the noise (\cref{CCC_Table}) \cite{PhysRevB.101.075308,SM}. Notably, in recent experiment 
auto- and cross-correlation data are indeed found to be 
\emph{unequal} \cite{DCG}.

\textit{The $G_{D_1}=5/9$ plateau.---} 
Here only the unequilibrated scenario is possible and similarly to the $1/2$ QPC conductance plateau, we can identify three case:
(A) We expect no QPC conductance plateau and no noise in the coherent regime, assuming clean contacts screening the interaction between the modes \cite{Protopopov2017, PhysRevB.104.115416} leading to a $2e^2/h$ two-terminal conductance for a fully open QPC. Considering stochastic scattering, noise
would arise even in the absence of QPC incompatible with the observations.
(B) If the $e/3, e/3, n_1, n_2$ modes, at the Wang-Meir-Gefen (WMG) RG fixed point, emanate from the 
contacts, undergoing coherent scattering at QPC, we find 
$G_{D_1}=5/9$ and no noise. 
(C) Here the $e/3, e/3, n_1, n_2$ modes emanate from the 
contacts and the stochastic (deterministic) incoherent scattering processes occur at regions
$\Tilde{D}_1, \Tilde{D}_2\ (\Tilde{S}, \Tilde{G})$
of size $l'_{\text{RG}}$ near the QPC (\cref{5by9QPC}),
We write down an infinite series
with contributions (i), (ii), and (iii) leading to
$\langle I \rangle = 2e/3\tau$, hence
$t = 5/6$, and $G_{D_1} = 5/9$ and $F_1 = F_2 = -F_c \approx 0.26$ (\cref{CCC_Table}). 

\textit{The $G_{D_1}=1/3$ plateau.---} 
Earlier experiments have shown the emergence of $G_{D_1}=1/3$ \cite{Cunningham1992,Mahalu2009,Bid2010} and here, we theoretically explain it based on either unequilibrated or equilibrated
scenarios and show that shot noise can be used to discriminate among
those (\cref{CCC_Table}).

(a) Unequilibrated scenario.--- 
A QPC plateau is observed at transmission $t=1/2$, leading to $G_{D_1}=1/3$ \cite{Mahalu2009} (\cref{1by3QPC}(a)). 
It was shown earlier that the neutral 
modes can create particle-hole
pairs, which stochastically split
and reach different drains, thus creating current 
fluctuations in $D_1,D_2$, leading to $F_1=F_2=2/3$ \cite{Sabo2017,Park2021}. Using the same stochastic variable approach, we find $F_c=-2/3$. We note that the coherent
regime remains noiseless.

(b) Equilibration scenario.--- 
We employ similar technique as for the $1/2$ QPC conductance plateau, thus, full charge equilibration leads to $G_{D_1} = 1/3, t=1/2$ (\cref{1by3QPC}(b))
for two possible edge 
structures (\cref{CCC_Table}).
For no thermal equilibration (one edge was analysed in Ref.\ \onlinecite{Kumar2022}) we have 
constant Fano factors. 
Let us also note that, in this case, Ref.\ \onlinecite{Park2024} pointed out a different situation where at the noise spot, not all the modes reach a common temperature, but rather the QPC mode(s) slowly give up heat to the bulk filling modes mode while travelling around the QPC, leading to length dependent noise.
Finally, and regardless of the previous statement, for hybrid and full
thermal equilibration, we find that the Fano factors acquire a $\sqrt{L_{\text{A}}/l_{\text{eq}}^{\text{th}}}$ contribution (\cref{CCC_Table}) \cite{PhysRevB.101.075308}.

\textit{Summary and outlook.---}
\SM{We have proposed a new paradigm for the emergence of unorthodox quantized conductance plateaus in the context of bulk fractional
quantum Hall QPC devices. It relies on the interaction of the bulk and QPC edge modes, which either coherently flow to new renormalized modes or incoherently equilibrate to differing degrees. Interestingly, distinct such scenarios may give rise to the same plateau; distinguishing between them requires the use of another unique feature, the inequality of auto- and cross-correlation noise \cite{DCG}.
As a prototypical example we analyzed the 2/3 state, uncovering several possible scenarios for the 
recently observed $e^2/2h$ \cite{Manfra2022,Hirayama2023} and  $e^2/3h$ \cite{Mahalu2009} plateaus and the means to distinguish between them via shot noise. We also predict a possible $5e^2/9h$ unequilibrated plateau. Our scheme is realizable with present day experimental facilities, and can be extended to other quantum Hall states \cite{SMAD}, graphene quantum Hall \cite{Pandey2024}, and edge reconstructed $\mathbb{Z}_2$ topological insulators \cite{PhysRevLett.128.186801, PhysRevLett.118.046801, John_2022}.}

\begin{acknowledgements}
We thank Christian Glattli, Udit Khanna,
Michael J.\ Manfra, Alexander D.\ Mirlin,
Jinhong Park, Christian 
Sp\aa{}nsl\"att, and Kun Yang for useful discussions.
We thank Christian Glattli also for sharing his unpublished data.
S.M.\ was supported by the Weizmann Institute of Science, Israel Deans fellowship through Feinberg
Graduate School, as well as the Raymond Beverly Sackler Center for Computational Molecular and Material Science at Tel Aviv University. 
A.D.\ was supported by the German-Israeli Foundation Grant No. I-1505-303.10/2019, DFG
MI 658/10-2, DFG RO 2247/11-1, DFG EG 96/13-1,
and CRC 183 (project C01). A.D.\ also thanks the Israel Planning and Budgeting Committee (PBC) and the
Weizmann Institute of Science, the Dean of Faculty fellowship, and the Koshland Foundation for financial support, as well as an IISER Tirupati start-up grant. Y.G.\ acknowledges support by
the DFG Grant MI 658/10-2., by grant no 2022391 from the
United States - Israel Binational Science Foundation
(BSF), and by the Minerva foundation. Y. G.\ is an incumbent of InfoSys chair at IISc.
M.G.\ has been supported by the Israel Science Foundation (ISF) and the Directorate for Defense
Research and Development (DDR\&D) Grant No. 3427/21, the ISF grant No. 1113/23, and the BSF Grant No. 2020072. 
\end{acknowledgements}

\bibliography{bibfile}

\begin{thebibliography}{58}
\bibitem [{ave()}]{averagedef}%
  \BibitemOpen
  \href@noop {} {}\bibinfo {note} {We define the time average of an observable
  $x$ as $\langle x \rangle = \lim_{\tau \rightarrow \infty} \frac{1}{\tau}
  \int_{-\tau}^{\tau} x(t) dt $}\BibitemShut {NoStop}%
\end{thebibliography}

\clearpage
\newpage
\onecolumngrid
\global\long\def\thesection{S\Alph{section}}
\global\long\def\thesubsection{\Roman{subsection}}
\setcounter{equation}{0}
\setcounter{figure}{0}
\setcounter{table}{0}
\setcounter{page}{1}
\renewcommand{\theequation}{S\arabic{equation}}
\renewcommand{\thefigure}{S\arabic{figure}}
\renewcommand{\bibnumfmt}[1]{[S#1]}
\renewcommand{\citenumfont}[1]{S#1}

\bigskip
\begin{center}
	\large{\bf Supplemental Material for ``Multiple Mechanisms for Emerging Conductance Plateaus in Fractional Quantum Hall States"}\\
\end{center}
\begin{center}
	Sourav Manna$^{1,2}$, Ankur Das$^{1,3}$, Yuval Gefen$^{1}$ and Moshe Goldstein$^{2}$
	\\
	{\it $^{1}$Department of Condensed Matter Physics, Weizmann Institute of Science, Rehovot 7610001, Israel\\$^{2}$Raymond and Beverly Sackler School of Physics and Astronomy, Tel-Aviv University, Tel Aviv, 6997801, Israel\\$^{3}$Department of Physics, Indian Institute of Science Education and Research (IISER) Tirupati, Tirupati 517619, India}\\
\end{center}

	The supplemental material contains the 
	definitions of the Fano factors (Sec.\ SI) and the details of the shot noise 
	on the $e^2/2h$ quantum point contact (QPC) conductance
	plateau (Sec.\ SII). The analysis for the other two QPC plateaus is similar.

\twocolumngrid

\section{SI. Definitions}\label{Fano}
We define the time dependent current-current correlations (CCC) $\delta^2 I_{ij}(\bar{t})$ as the average of the symmetric combination of the product of two current operators \cite{MartinNoiseReview}
\begin{equation}
	\begin{split}
		\delta^2 I_{ij}(\bar{t}) &= \langle \{ \Delta I_i(\bar{t}), \Delta I_j(0)\} \rangle \\& = \langle \Delta I_i(\bar{t}) \Delta I_j(0) \rangle + \langle \Delta I_j(0) \Delta I_i(\bar{t}) \rangle, 
	\end{split}
\end{equation}
where for $i=j$, $\delta^2 I_{ii}(\bar{t})$ is referred to as an autocorrelation, and for $i \neq j$, $\delta^2 I_{ij}(\bar{t})$ is referred to as a crosscorrelation. Here, $\Delta I_i(\bar{t}) = (I_i(\bar{t}) - \langle I \rangle)$ is for the $i$th drain.
We note that the product of two current operators evaluated at different times is not Hermitian, and hence we define $\delta^2 I_{ij}(\bar{t})$ using an anti-commutator. 
In the frequency domain we write 
\begin{equation}
	\begin{split}
		\delta^2 I_{ij} (\omega) &= \int_{-\infty}^{\infty} \delta^2 I_{ij}(\bar{t}) e^{i w \bar{t}} d\bar{t},
	\end{split}
\end{equation}
where $\omega$ is the frequency. In the dc limit or zero frequency limit we have
\begin{equation}
	\begin{split}
		\delta^2 I_{ij} (\omega \rightarrow 0) =\frac{2 \delta^2 Q_{ij}}{\tau^2},
	\end{split}
\end{equation}
with charge at the $i$th drain being $Q_i=\lim_{\tau \rightarrow \infty}\int_{0}^{\tau}I_i(t)dt$, $\tau$ is the time,
and $\delta^2 Q_{ij}$ is the 
correlation in charge fluctuation. With the source current denoted by $I$ and transmission coefficient denoted by $t$, the Fano factor is defined as
\begin{equation}
	\begin{split}
		F &= \frac{\delta^2 I_{ij} (\omega \rightarrow 0)}{2 (\frac{e}{\tau}) I t (1-t)}=\frac{\delta^2 Q_{ij}}{ e \tau I t (1-t)}.
	\end{split}
\end{equation}

\section{SII. The \texorpdfstring{$e^2/2h$}{e2h} plateau}\label{1by2QPC_PRB}

Here, we show how $e^2/2h$ QPC conductance plateau 
(experimentally discovered in Ref.\ \cite{Manfra2022,Hirayama2023},
following an earlier theoretical work in Ref.\ \cite{PhysRevB.87.125130} (see also Refs.\ \onlinecite{Wang2021, Yuli2023}))
can appear
both in the unequilibrated and equilibrated regimes. We compute
the shot noise at this plateau and show that different inequalities 
hold among the autocorrelations and crosscorrelation
for different models. In the unequilibrated regime, we consider both the
coherent and stochastic scenarios.

\subsection{Coherent scenario}

We start with the MacDonald edge containing the 
$e/3$ and $e$
charge modes (counter-propagating), while going 
towards the edge from bulk
\cite{PhysRevLett.64.220} (\cref{KFP_Coherent_Ideal_Fig,KFP_Coherent_NonIdeal_Fig,KFP_Fig}).
Let us label the charge $e$ mode as mode $``1"$ and the charge $e/3$ mode as mode $``2"$. 
In each region
between a contact and QPC the $e$ and $e/3$ modes 
are renormalized at the Kane-Fischer-Polchinski (KFP)
renormalization group (RG) fixed point leading to 
counter-propagating $2e/3$ charge mode and $n$ neutral mode (KFP region) \cite{Kane1994}.
At the QPC the $e/3$ mode is fully backscattered and the $e$ mode is fully transmitted.
We will now consider two cases: ideal contact and
non-ideal contact. We stress that for the ideal contact, we
\emph{do not} have a QPC conductance plateau. However, 
for the non-ideal
contact we have a transmission $t$, leading to a $G_{D_1}e^2/h$ QPC conductance plateau, where $G_{D_1}=t I \tau/e$ and $I$ is the source current.

\subsubsection{Ideal contact}

\begin{figure}[!t]
	\includegraphics[width=\columnwidth]{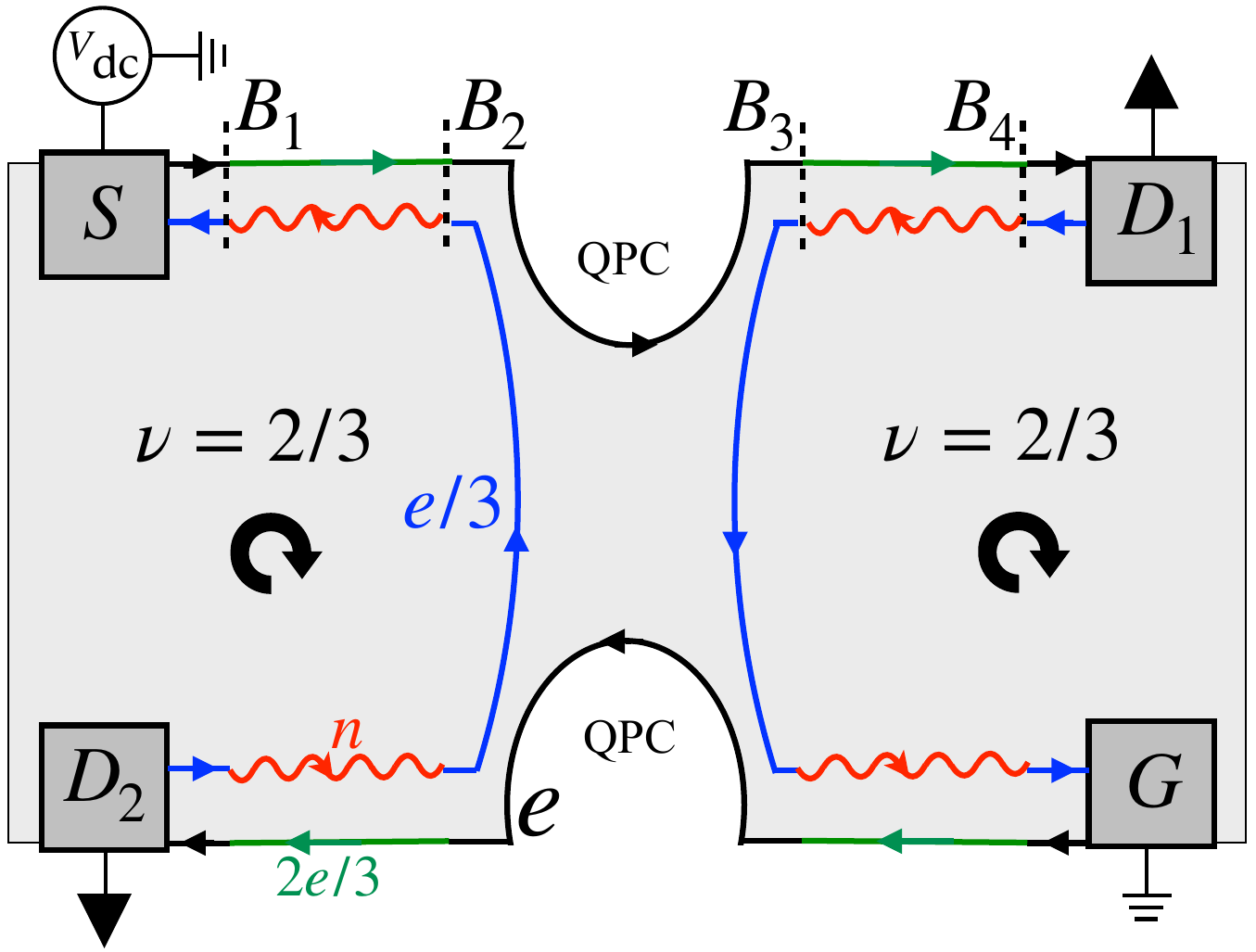}
	\caption{The coherent scenario with clean 
		contacts where the modes do not interact \cite{Protopopov2017, PhysRevB.104.115416}: We show a Hall bar at filling $\nu=2/3$ constricted by a QPC. We have four contacts: a source
		$S$ biased by a dc voltage $V_{\text{dc}}$, a ground contact $G$, and two drains $D_1$ and $D_2$. Each circular arrow shows the chirality of charge propagation. 
		While going from bulk towards the edge, the $e/3$ and $e$ charge modes counter-propagate (MacDonald edge) \cite{PhysRevLett.64.220}. These modes are renormalized
		to a $2e/3$ charge mode and a 
		counter-propagating $n$ neutral mode at the 
		KFP RG fixed point. At the QPC, full backscattering occurs
		for the $e/3$ mode, while the $e$ mode is transmitted
		fully. $B_1, B_2, B_3, B_4$ denote the boundaries
		between non-interacting modes and KFP regions, which
		appear in the arms, while going from $S$ to $D_1$.}\label{KFP_Coherent_Ideal_Fig}
\end{figure}

Let us first assume that contacts are clean and the modes do not
interact therein \cite{Protopopov2017, PhysRevB.104.115416} (\cref{KFP_Coherent_Ideal_Fig}).
Starting from the source $S$ at charge mode $e$, a 
wavepacket encounters an infinite number of 
transmissions and reflections while entering (through $B_1$) and
leaving (through $B_2$) the KFP region in the arm between $S$ and QPC.
We calculate the net transmitted charge ($q$),
coming out of $S$, by summing up an infinite series as 
\begin{equation}
	\begin{split}
		q &= e T_{11}^{B_1}T_{11}^{B_2} \Big[ 1+ R_{12}^{B_2}R_{21}^{B_1}+ \dots \Big]\\&
		=e\Bigg(  \frac{T_{11}^{B_1}T_{11}^{B_2}}{1-R_{12}^{B_2}R_{21}^{B_1}} \Bigg),
	\end{split}
\end{equation}
where $T$ and $R$ denote the 
transmission and reflection coefficients respectively.
Following Ref.\ \onlinecite{Protopopov2017}, we use
$T=\sqrt{2/3}, R=\sqrt{1/3}$ and find $q=e$.

This fully-transmitted charge $q$ arrives (via the charge mode $e$) at the arm
between QPC and $D_1$ and enters the KFP region there
via $B_3$. 
Then it encounters an infinite number of 
transmissions and reflections while entering (through $B_3$) and
leaving (through $B_4$) the KFP region in the arm between QPC and $D_1$.
We calculate the net reflected charge ($q'$) 
from $B_3$ by summing up an infinite series as 
\begin{equation}
	\begin{split}
		q' &= e \Big[ R_{12}^{B_3}+ T_{11}^{B_3}R_{12}^{B_4}T_{22}^{B_3} \Big(1+R_{21}^{B_3}R_{12}^{B_4}+ \dots\Big) \Big]\\&
		=e\Bigg( R_{12}^{B_3}+ \frac{T_{11}^{B_3}R_{12}^{B_4}T_{22}^{B_3}}{1-R_{21}^{B_3}R_{12}^{B_4}} \Bigg).
	\end{split}
\end{equation}
Following Ref.\ \onlinecite{Protopopov2017}, we use
$R_{12}^{B_3}=-\sqrt{1/3}, T=\sqrt{2/3}$ and 
all other $R=\sqrt{1/3}$ and find $q'=0$.
Therefore, all the charges from $S$ reach $D_1$ and
there is no QPC conductance plateau. Also, there is no
stochastic process involved here and hence no shot noise. We note that this scenario leads to a $4e^2/3h$ two-terminal conductance for a fully open QPC.

Let us mention another possibility, where one considers incoherent stochastic scattering in the transition regions between the different segments (see \cref{KFP_stoch} below). In that case
noise would be present even without the QPC, which is excluded by the observations.

\subsubsection{Non-ideal contact}

\begin{figure}[!t]
	\includegraphics[width=\columnwidth]{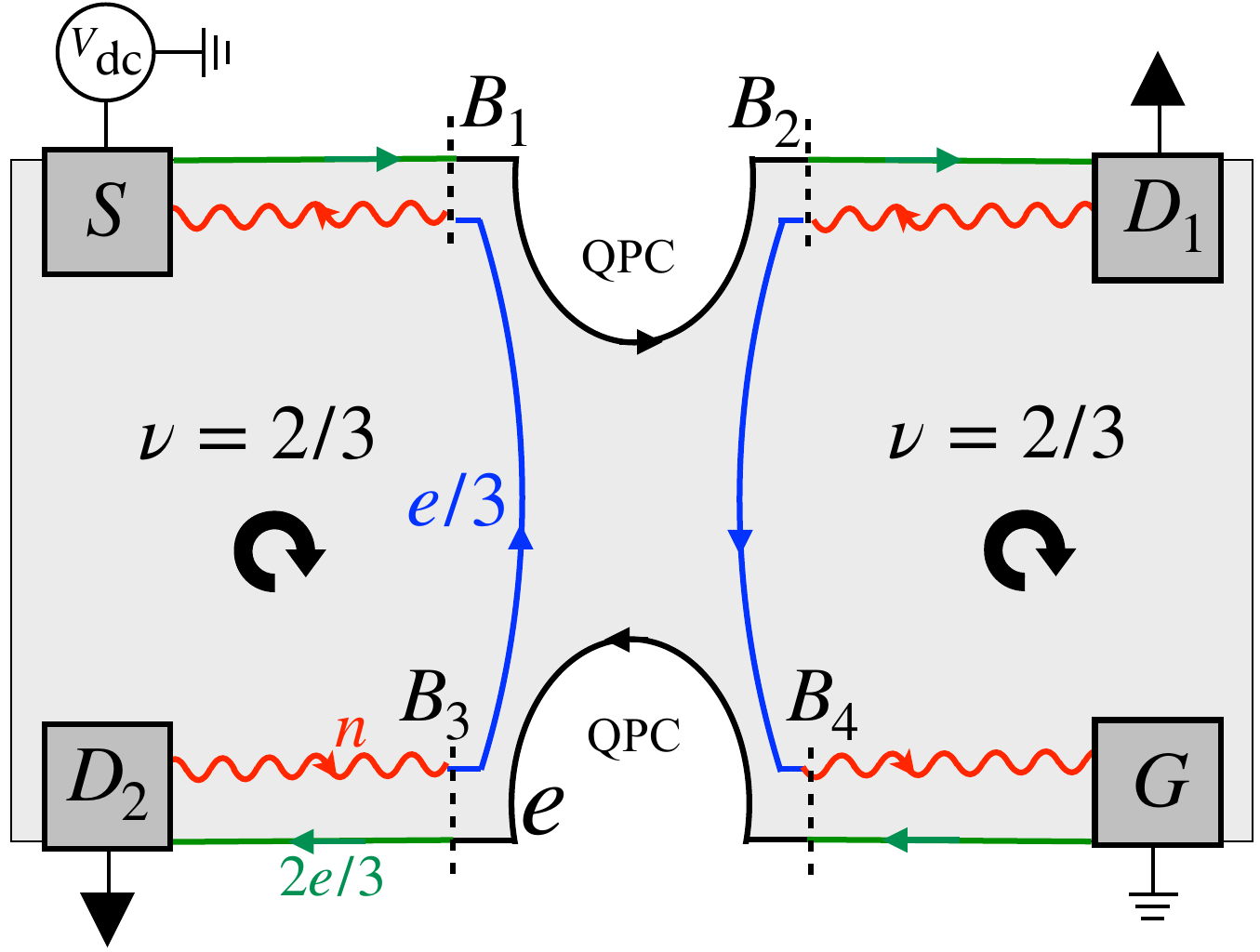}
	\caption{The coherent scenario for the $1/2$ QPC conductance plateau where the
		renormalized modes emanate from the contacts: The set up is similar to \cref{KFP_Coherent_Ideal_Fig}. $B_1, B_2, B_3, B_4$ denote the boundaries
		between non-interacting modes and KFP regions near QPC.}\label{KFP_Coherent_NonIdeal_Fig}
\end{figure}

We shall assume that the renormalized modes emerge from the
contacts (\cref{KFP_Coherent_NonIdeal_Fig}). 
In a similar manner as for the ideal contact case, we
consider that a wavepacket from $S$ encounters
an infinite number of 
transmissions and reflections at boundaries 
$B_1, B_2, B_3, B_4$.
We calculate the charge reaching $D_1$ by summing up an infinite series as 
\begin{equation}
	\begin{split}
		Q_1 &= \frac{2e}{3} T_{11}^{B_1}T_{11}^{B_2} \Big[ 1+ R_{12}^{B_2}R_{21}^{B_4}R_{12}^{B_3}R_{21}^{B_1}+ \dots \Big]\\&
		=\frac{2e}{3}\Bigg(  \frac{T_{11}^{B_1}T_{11}^{B_2}}{1-R_{12}^{B_2}R_{21}^{B_4}R_{12}^{B_3}R_{21}^{B_1}} \Bigg).
	\end{split}
\end{equation}
Following Ref.\ \onlinecite{Protopopov2017}, we use
$T=\sqrt{2/3}, R=\sqrt{1/3}$ and find $Q_1=e/2$.
The source current is $I=2e/3$ leading to a
QPC transmission $t=Q_1/I=3/4$, and this gives rise to a QPC conductance plateau at $G_{D_1}=1/2$.
We note that no
stochastic process is involved here and hence no shot noise.

\subsection{Stochastic scenario}\label{KFP_stoch}

\begin{figure}[!t]
	\includegraphics[width=\columnwidth]{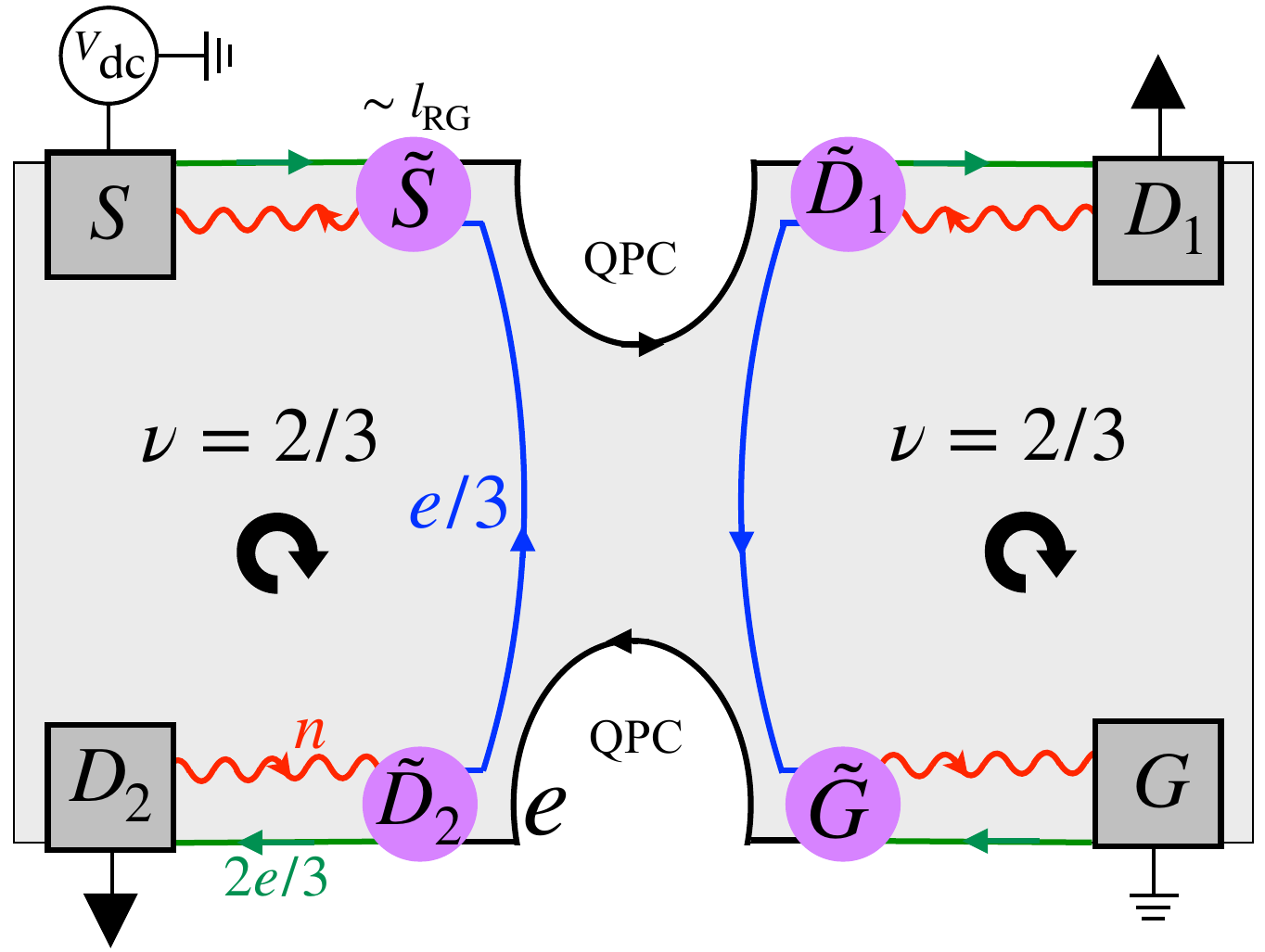}
	\caption{The stochastic scenario for the $1/2$ QPC conductance plateau where the
		renormalized modes emanate from the contacts: The set up is similar to \cref{KFP_Coherent_Ideal_Fig}.
		Near the QPC, we have scattering regions of size $l_{\text{RG}}$ (purple circles) for which we define by density kernel matrices $S_{\Tilde{S}}, S_{\Tilde{D}_1}, S_{\Tilde{G}}, S_{\Tilde{D}_2}$ \cite{PhysRevB.52.R17040} 
		constructed out of the
		transmission and reflection coefficients. We note that
		$S_{\Tilde{D}_1}$ and $S_{\Tilde{D}_2}$ give rise to the stochastic processes
		while at $S_{\Tilde{S}}$ and $S_{\Tilde{G}}$ deterministic charge forward scattering occurs.}\label{KFP_Fig}
\end{figure}

Finally, we will assume that the renormalized modes emerge from the
contacts (\cref{KFP_Fig}), and furthermore, that we have incoherent scattering of wavepackets in
each region of size $l_{\text{RG}}$ near QPC.
We parametrize these processes by the density kernel matrices 
$S_{\Tilde{S}}, S_{\Tilde{D}_1}, S_{\Tilde{G}}, S_{\Tilde{D}_2}$ \cite{PhysRevB.52.R17040}. 
We point out that $S_{\Tilde{D}_1}$ and $S_{\Tilde{D}_2}$ are responsible for stochastic processes where incoming charge is split between the outgoing modes,
while $S_{\Tilde{S}}$ and $S_{\Tilde{G}}$ describe deterministic processes, where incoming charge has only a single outgoing mode to proceed to.
Explicitly, we write
\begin{equation}
	\begin{split}
		&S_{\Tilde{S}}=
		\begin{pmatrix}
			T_{11}  & R_{21} \\
			R_{12} &  T_{22} 
		\end{pmatrix};
		S_{\Tilde{D}_1}=
		\begin{pmatrix}
			\langle T'_{11} \rangle & \langle R'_{21} \rangle\\
			\langle R'_{12} \rangle & \langle T'_{22} \rangle 
		\end{pmatrix},\\&
		S_{\Tilde{G}}=
		\begin{pmatrix}
			T''_{11}  & R''_{21} \\
			R''_{12} &  T''_{22} 
		\end{pmatrix};
		S_{\Tilde{D}_2}=
		\begin{pmatrix}
			\langle T'''_{11} \rangle & \langle R'''_{21} \rangle\\
			\langle R'''_{12} \rangle & \langle T'''_{22} \rangle 
		\end{pmatrix},
	\end{split}
\end{equation}
where $T',T''',R',R'''$ are Bernouli random numbers $\in \{0,1\}$ which take account for the stochastic tunneling processes and $T,T'',R,R''$ are deterministic.
We denote by $\langle \cdots \rangle$ the first moment (average value) of
the distribution of these stochastic variables.

We calculate the total charge, which reach different contacts via multiple reflections and transmissions
throughout. Thereby, we sum up the resulting infinite series.
We organize this infinite series into three different factors:
\begin{enumerate}
	\item A factor accounting for the first tunnelling while entering the QPC.
	\item The shortest route inside the QPC.
	\item Different scatterers give rise to the contribution due to multiple reflections. This factor $(R'_{12})_i (R''_{21})_i (R'''_{12})_i (R_{21})_{i+1},\ i \in [1,2,\ldots]$ will remain same for all the contacts.
\end{enumerate}
For the charge $Q_1$ entering at drain $D_1$ the first contribution is $(T_{11})_1$,
and the second contribution is $(T'_{11})_n,\ n\in[1,2,\ldots]$. The number outside of the parenthesis denotes how
may times the wavepacket visited the same scatterer. Thus, using this we can write $Q_1$ as
\begin{widetext}
	\begin{equation}
		\begin{split}
			Q_1 &= \frac{2e}{3} \Big[ (T_{11})_1 (T'_{11})_1 + (T_{11})_1  (R'_{12})_1 (R''_{21})_1 (R'''_{12})_1 (R_{21})_2  (T'_{11})_2 \\&+ (T_{11})_1  (R'_{12})_1 (R''_{21})_1 (R'''_{12})_1 (R_{21})_2 (R'_{12})_2 (R''_{21})_2 
			(R'''_{12})_2 (R_{21})_3 (T'_{11})_3 \\&+ (T_{11})_1  (R'_{12})_1 (R''_{21})_1 (R'''_{12})_1 (R_{21})_2 (R'_{12})_2 (R''_{21})_2 (R'''_{12})_2 (R_{21})_3  (R'_{12})_3 (R''_{21})_3 (R'''_{12})_3 (R_{21})_4  (T'_{11})_4 + \ldots  \Big].
		\end{split}
	\end{equation}
\end{widetext}

We note the following rules,
\begin{equation}\label{Eq-relation}
	\begin{split}
		&(T'_{11})_i + (R'_{12})_i = 1,\ (T'_{11})_i (R'_{12})_i = 0,\\&(T'_{22})_i + (R'_{21})_i = 1,\ (T'_{22})_i (R'_{21})_i = 0,
	\end{split}
\end{equation}
and for different scattering event indices $i$, any $T'$ and $R'$ are uncorrelated. Similar relations hold for
$T''',R'''$. We also note that transmission and
reflection coefficients with different ``prime'' indices are uncorrelated as
those describe different scatterers. Therefore,
\begin{equation}\label{Eq-relation}
	\begin{split}
		&\langle Q_1 \rangle = \frac{2e}{3} \Big[  \frac{T_{11} \langle T'_{11} \rangle}{1-\langle R'_{12} \rangle R''_{21} \langle  R'''_{12} \rangle R_{21} } \Big].
	\end{split}
\end{equation}
We use $T_{11} =1, \langle T'_{11} \rangle=2/3, \langle R'_{12} \rangle=\langle R'''_{12} \rangle=1/3, R''_{21} = R_{21} =1$
to find $\langle Q_1 \rangle =e/2$. 
The source current is $I=2e/3\tau$ and the 
transmission coefficient is then 
$t=\langle Q_1 \rangle/\tau I=3/4$ and the 
QPC conductance plateau is at 
$G_{D_1}=1/2$.
The charge reaching $D_2$ is 
$\langle Q_2 \rangle = 2e/3 - \langle Q_1 \rangle=e/6$.

For the autocorrelation $\delta^2 Q_1=\langle Q^2_1 \rangle- \langle Q_1 \rangle^2 = \langle Q_1 \rangle(2e/3-\langle Q_1 \rangle)$,
since $\langle Q_1^2 \rangle = 2e \langle Q_1 \rangle/3$, at $D_1$ we obtain
$\delta^2 Q_1=e^2/12$.
Similarly, for the autocorrelation $\delta^2 Q_2=\langle Q^2_2 \rangle- \langle Q_2 \rangle^2$ at $D_2$ we obtain
$\delta^2 Q_2=e^2/12$
and the crosscorrelation $\delta^2 Q_c= \langle Q_1 Q_2 \rangle - \langle Q_1 \rangle \langle Q_2 \rangle = - \langle Q_1 \rangle \langle Q_2 \rangle$
(since $\langle Q_1 Q_2 \rangle = 0$) becomes
$\delta^2 Q_c=-e^2/12$.

The Fano factors are
\begin{equation}
	F_i = \frac{\delta^2 Q_i}{e \tau I t (1-t)},\ i \in \{1,2,c\},
\end{equation}
and therefore we find them
to be $F_1= F_2 = -F_c = 2/3$.

\subsection{Equilibration scenario}
\begin{figure}[!t]
	\includegraphics[width=\columnwidth]{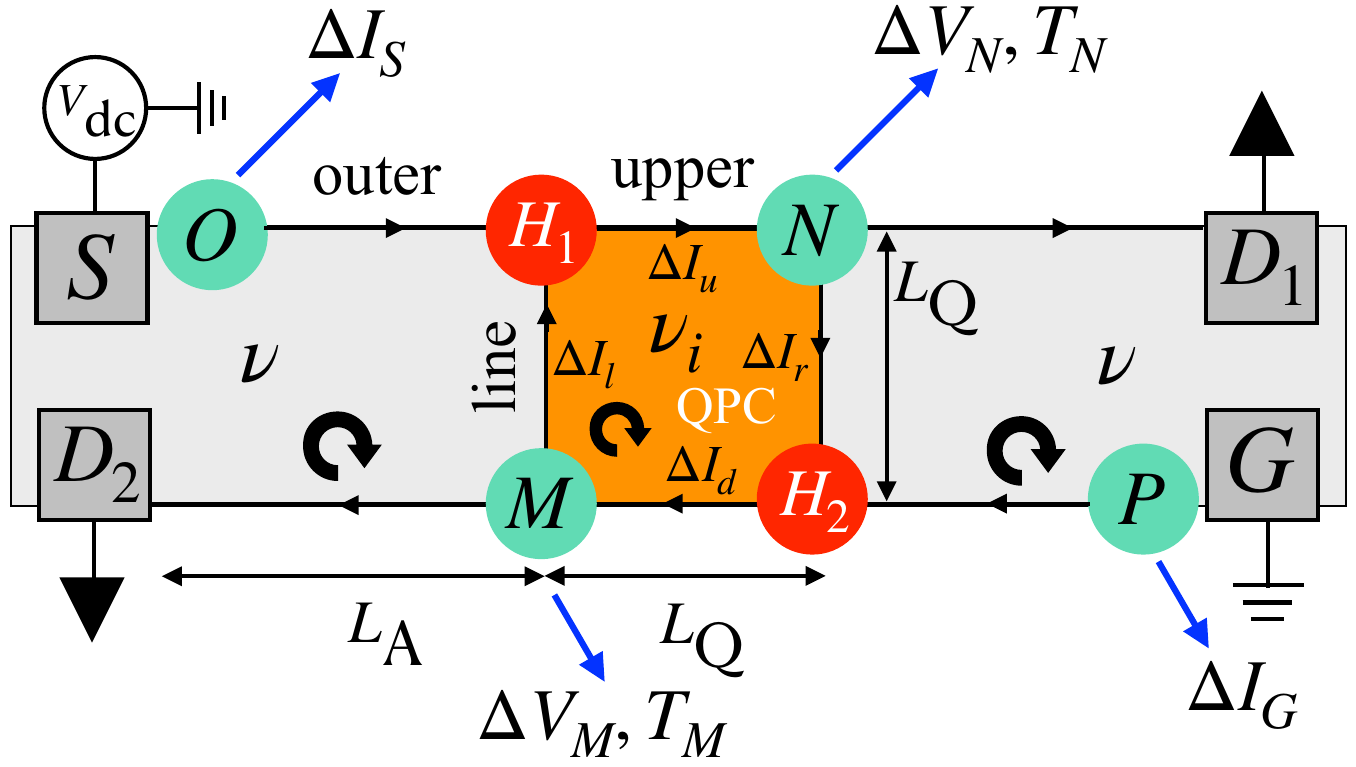}
	\caption{The equilibration scenario for the $1/2$ QPC conductance plateau: We show a Hall bar at filling $\nu$ with a QPC at filling $\nu_i$, where $\nu<\nu_i$. We have a source
		contact $S$ biased by a dc voltage $V_{\text{dc}}$, a ground contact $G$, and two drains $D_1$ and $D_2$. Each circular arrow shows the chirality of charge propagation. The geometric lengths are $L_{\text{A}}, L_{\text{Q}}$. The nomenclature we use:
		``outer" is the segment between the vacuum and $\nu$, 
		``upper" is the segment between the vacuum and $\nu_i$,
		and ``line" is the segment between $\nu$ and $\nu_i$.
		Here, the chirality of charge propagation of a given filling is shown by a circular arrow. We also indicate the noise generation in the setup:  
		$H_1,H_2$ are the hot spots (red circles) owing to the voltage drops and $M,N,O,P$ are the noise spots (green circles) \cite{PhysRevB.101.075308,SM5by2}.} \label{nuSmall_nuiLarge_SM}
\end{figure}

We derive the general expressions for the CCC (shot noise) in a QPC for the
bulk filling $\nu$ and QPC filling $\nu_i$ when $\nu<\nu_i$ (\cref{nuSmall_nuiLarge_SM}). We assume that the charge is fully equilibrated, hence charge transport is ballistic, moving ``downstream" along each segment of the setup.
We call the direction opposite to charge flow (``upstream") antiballistic.
Thereafter, we compute the values of CCC for 
specific choices of $\{\nu,\nu_i\}$ and for different
thermal equilibration regimes.

We consult \cref{nuSmall_nuiLarge_SM} and follow Refs.\ \cite{PhysRevB.101.075308,SM5by2} to write 
\begin{equation}\label{CurFluc}
	\begin{split}
		&\Delta I_S+\Delta I_l=\Delta I_u,\\&
		\Delta I_u=\Delta I_1+\Delta I_r,\\&
		\Delta I_G+\Delta I_r=\Delta I_d,\\&
		\Delta I_d=\Delta I_2+\Delta I_l,
	\end{split}
\end{equation}
where $\Delta I_i, i \in \{S, G, u, d, r, l, 1, 2\}$ are the current fluctuations. We also write
\begin{equation}\label{VolFluc}
	\begin{split}
		&\Delta I_1 = \nu \frac{e^2}{h}\Delta V_N + \Delta I_1^{\text{th}},\\&
		\Delta I_r = (\nu_i-\nu) \frac{e^2}{h}\Delta V_N + \Delta I_r^{\text{th}},\\&
		\Delta I_2 = \nu \frac{e^2}{h}\Delta V_M + \Delta I_2^{\text{th}},\\&
		\Delta I_l = (\nu_i-\nu) \frac{e^2}{h}\Delta V_M + \Delta I_l^{\text{th}},
	\end{split}
\end{equation}
where $\Delta V_i, i \in \{M,N\}$ are the voltage fluctuations and 
$\Delta I_i^{\text{th}}, i \in \{1,2,r,l\}$
are the thermal fluctuations. We find
\begin{equation}\label{D1Fluc}
	\begin{split}
		&\Delta I_1 = \frac{1}{(\nu-2\nu_i)}\Big[ (\nu-\nu_i)
		(\Delta I_G + \Delta I_1^{\text{th}} - \Delta I_2^{\text{th}})\\&\qquad
		- \nu (\Delta I_l^{\text{th}} - \Delta I_r^{\text{th}}) -\nu_i \Delta I_S \Big],\\&
		\Delta I_2 = \frac{1}{(\nu-2\nu_i)}\Big[ (\nu-\nu_i)
		(\Delta I_S - \Delta I_1^{\text{th}} + \Delta I_2^{\text{th}})\\&\qquad
		+ \nu (\Delta I_l^{\text{th}} - \Delta I_r^{\text{th}}) -\nu_i \Delta I_G \Big].
	\end{split}
\end{equation}
We use the local Johnson-Nyquist relations for thermal noise,
\begin{equation}
	\begin{split}
		&\ \ \ \ \  \langle (\Delta I_l^{\text{th}})^2 \rangle = \frac{2 e^2}{h} (\nu_i-\nu) k_{\text{B}} T_M,\\& \ \ \ \ \ \langle (\Delta I_1^{\text{th}})^2 \rangle = \frac{2 e^2}{h} \nu k_{\text{B}} T_N,\\& \ \ \ \ \ \langle (\Delta I_r^{\text{th}})^2 \rangle = \frac{2 e^2}{h}
		(\nu_i-\nu)k_{\text{B}} T_N,\\& \ \ \ \ \ \langle (\Delta I_2^{\text{th}})^2 \rangle = \frac{2 e^2}{h} \nu k_{\text{B}} T_M,\\& \langle (\Delta I_i^{\text{th}} \Delta I_j^{\text{th}}) \rangle = 0,\ \text{for}\ i \neq j\ \text{and}\ i,j \in \{1,2,l,r\},
	\end{split}
\end{equation}
where $k_{\text{B}}$ is the Boltzmann constant to write 
\begin{equation}
	\begin{split}
		\delta^2 I_1 &=2\Bigg(\frac{e^2}{h}\Bigg)\frac{\nu \nu_i (\nu_i-\nu)}{(\nu-2\nu_i)^2}k_{\text{B}} (T_M+T_N) \\&+ \frac{1}{(\nu-2\nu_i)^2}\Big[\nu_i^2 \langle (\Delta I_S)^2 \rangle + (\nu-\nu_i )^2\langle (\Delta I_G)^2 \rangle\Big],
	\end{split}
\end{equation}
\begin{equation}
	\begin{split}
		\delta^2 I_2 &=2\Bigg(\frac{e^2}{h}\Bigg)\frac{\nu \nu_i (\nu_i-\nu)}{(\nu-2\nu_i)^2}k_{\text{B}} (T_M+T_N) \\&+ \frac{1}{(\nu-2\nu_i)^2}\Big[\nu_i^2 \langle (\Delta I_G)^2 \rangle + (\nu-\nu_i )^2\langle (\Delta I_S)^2 \rangle\Big],
	\end{split}
\end{equation}
and
\begin{equation}
	\begin{split}
		\delta^2 I_c &=-2\Bigg(\frac{e^2}{h}\Bigg)\frac{\nu \nu_i (\nu_i-\nu)}{(\nu-2\nu_i)^2}k_{\text{B}} (T_M+T_N) \\&+ \frac{\nu_i(\nu_i-\nu)}{(\nu-2\nu_i)^2}\Big[\langle (\Delta I_S)^2 \rangle + \langle (\Delta I_G)^2 \rangle\Big],
	\end{split}
\end{equation}
where $T_M, T_N$ are, respectively, the temperatures at the noise spots $M$ and $N$ which are found by solving
self-consistent equilibration equations and 
considering energy conservations \cite{SM5by2,PhysRevB.101.075308}. We note that the 
dissipated powers at the hot spots take the form
\begin{equation}
	\begin{split}
		P_{H_1} = P_{H_2} = \frac{e^2 V_{\text{dc}}^2}{h}\frac{\nu(\nu_i-\nu)}{2\nu_i} \Big( \frac{\nu_i}{2 \nu_i - \nu}  \Big)^2.
	\end{split}
\end{equation}
The contributions
$\langle (\Delta I_G)^2 \rangle = \langle (\Delta I_S)^2$
at the noise spots $O$ and $P$ are computed by evaluating an integral as shown in the 
main text (also in Ref.\ \onlinecite{SM5by2,PhysRevB.101.075308})
assuming that the lead contacts remain at zero temperature.
We note that the Joule heating term (owing to the
voltage drop at the hot spots) 
should be a negligible contribution to those noise spots \cite{SM5by2,PhysRevB.101.075308}.

We refer to \cref{nuSmall_nuiLarge_SM} and consider $\{ \nu,\nu_i \} = \{ 2/3,1 \}, \{ 2/3(\text{r}),1 \},$ and $ \{ 2/3(\text{r}), 1(\text{r}) \}$,
where $2/3(\text{r})$ refers to the reconstructed MacDonald edge \cite{PhysRevLett.72.2624, Wang2013} leading to the filling factor discontinuity $\delta \nu = [-1/3,+1, -1/3, +1/3]$ and
$1(\text{r})$ denotes edge reconstruction in QPC leading to the QPC filling factor discontinuity
(from bulk to edge) $\delta \nu_i = [+1, -1/3, +1/3]$ \cite{Khanna2021}.
As charge is fully equilibrated in each segment of the QPC set up (\cref{nuSmall_nuiLarge_SM}), we have 
\begin{equation}\label{I1}
	\begin{split}
		I_1 = \frac{e^2 V_{\text{dc}}}{h} \times \frac{2}{3}\times \frac{2}{3}\times \sum_{i=0}^{\infty}\Big( \frac{1}{3^2} \Big)^i=\frac{e^2 V_{\text{dc}}}{2h}
	\end{split}
\end{equation}
for each of the $\{ \nu,\nu_i \}$ choices here, leading to the 
transmission $t=3/4$ and $G_{D_1}=1/2$. 

We consider three different regimes of thermal equilibration as 
while each segment of the QPC geometry is thermally
unequilibrated leading to
$L_{\text{Q}} \ll L_{\text{A}} \ll l_{\text{eq}}^{\text{th}}$ (no thermal equilibration),
only QPC segment of the geometry is thermally
unequilibrated and other 
segments are thermally equilibrated leading to
$L_{\text{Q}} \ll l_{\text{eq}}^{\text{th}} \ll L_{\text{A}}$ (hybrid thermal equilibration),
and each segment of the QPC geometry is thermally equilibrated leading to
$l_{\text{eq}}^{\text{th}} \ll L_{\text{Q}} \ll L_{\text{A}}$ (full thermal equilibration).
Here, $l_{\text{eq}}^{\text{th}}$
is the thermal equilibration length. We note that the thermally unequilibrated (but electrically equilibrated) scenario was also considered for a single edge segment in Ref.\ \onlinecite{Kumar2022}.
In the no thermal equilibration regime, only the upstream modes from
$H_1$($H_2$) will contribute to $O$($P$);
the downstream modes will not contribute as they have zero temperature (which is the contact temperature)
and are fully thermally
isolated from the upstream modes.

For no thermal equilibration, we have only ballistic and
antiballistic heat transports in any segment of the QPC set up. Our analyses are based on the
assumption that each of the modes creates noise
by producing particle-hole pairs at the noise spots
due to very short charge equilibration length. These processes lead to local heat
equilibration by transfering heat
energy among the modes. Thereby, all the modes
acquire a common temperature leading to constant CCC. 

Following
our assumptions, below we derive explicitly the CCC values for
$\{ \nu,\nu_i \} = \{ 2/3,1 \}$.
Conservation of energy at $H_1, M$ leads to the
following expressions \cite{SM5by2,PhysRevB.101.075308} :
\begin{equation}
	\begin{split}
		&J_S+J_l+P_{H_1}=J_u,\\&J_d=J_l+J_2,
	\end{split}
\end{equation}
where $J_i$ is the heat current along the segment
$i\in\{S, l, u, d, 2\}$. We write
\begin{equation}
	\begin{split}
		&J_S=-\frac{\kappa}{2}T^2_{H_1}, J_l=\frac{\kappa}{2}(2T^2_M-T^2_{H_1}),\\&
		J_u=\frac{\kappa}{2}T^2_{H_1},J_d=\frac{\kappa}{2}T^2_{H_2},J_2=\frac{\kappa}{2}T^2_{M},
	\end{split}
\end{equation}
where $\kappa=\frac{\pi^2 k^2_{\text{B}}}{3h}$ and 
$T_{H_1}=T_{H_2}$ and $T_M=T_N$ are the temperatures at $H_1, H_2, M, N$ respectively. We find 
\begin{equation}
	\begin{split}
		&T_M=T_N=\frac{3\sqrt{4} eV_{\text{dc}}}{4\sqrt{15} \pi k_{\text{B}}},
		T_{H_1}=T_{H_2}=\frac{3\sqrt{2} eV_{\text{dc}}}{4\sqrt{5} \pi k_{\text{B}}}.
	\end{split}
\end{equation}
The contribution from $O$ and $P$ to CCC is evaluated
to \cite{SM5by2,PhysRevB.101.075308}
\begin{equation}
	\begin{split}
		\langle (\Delta I_S)^2 \rangle = \langle (\Delta I_G)^2 \rangle \approx 0.0472 \frac{e^3 V_{\text{dc}}}{h}
	\end{split}
\end{equation}
leading to $F_1=F_2\approx 0.36, F_c \approx -0.17$.

In a similar manner, for 
$\{ \nu,\nu_i \} = \{ 2/3(\text{r}),1 \}$ we find 
\begin{equation}
	\begin{split}
		&T_M=T_N=\frac{ 3eV_{\text{dc}}}{16 \pi k_{\text{B}}},\\&\langle (\Delta I_S)^2 \rangle = \langle (\Delta I_G)^2 \rangle \approx 0.048 \frac{e^3 V_{\text{dc}}}{h},
	\end{split}
\end{equation}
leading to $F_1=F_2\approx 0.23, F_c \approx -0.04$.
For 
$\{ \nu,\nu_i \} = \{ 2/3(\text{r}),1(\text{r}) \}$ we find 
\begin{equation}
	\begin{split}
		&T_M=T_N=\frac{3\sqrt{5} eV_{\text{dc}}}{4\sqrt{36} \pi k_{\text{B}}},\\&\langle (\Delta I_S)^2 \rangle = \langle (\Delta I_G)^2 \rangle \approx 0.065 \frac{e^3 V_{\text{dc}}}{h},
	\end{split}
\end{equation}
leading to $F_1=F_2\approx 0.34$ and $F_c \approx -0.08$.
Similarly, for hybrid and full thermal equilibration, we have diffusive 
heat transport in the outer segment and 
ballistic and
antiballistic heat transports in the line and
the upper segments of the QPC set up leading to length
dependent CCC. For each choice of $\{ \nu,\nu_i \}$, we find
\begin{equation}
	\begin{split}
		&F_1=F_2\approx 0.13+0.26\sqrt{L_{\text{A}}/l_{\text{eq}}^{\text{th}}},\\&
		F_c\approx 0.08-0.26\sqrt{L_{\text{A}}/l_{\text{eq}}^{\text{th}}}.
	\end{split}
\end{equation}

\end{document}